\documentclass[10 pt, conference]{IEEEtran}
\usepackage{graphicx} % Required for inserting images
\usepackage{enumitem}
\usepackage{amsmath}
\usepackage{soul}
\usepackage{circledsteps}
\usepackage{fontawesome5}
\usepackage{tcolorbox}
\usepackage{array}
\usepackage{booktabs}
\usepackage{MnSymbol}
\usepackage{balance}

\newcolumntype{P}[1]{>{\centering\arraybackslash}p{#1}}

\newcommand{\question}[1]{%
    \medskip%
    \noindent\fcolorbox{black}{blue!05}{%
        \parbox{0.97\linewidth}{% 
            \textbf{\faQuestionCircle\;RQ} #1%
        }%
    }%
    \medskip%
}%

\newcommand{\answer}[1]{%
    \medskip%
    \noindent\fcolorbox{black}{green!05}{%
        \parbox{0.97\linewidth}{% 
            \textbf{\faExclamationCircle\;RQ} #1%
        }%
    }%
    \medskip%
}%

\begin{document}

\title{Designing Secure AI-based Systems: \\a Multi-Vocal Literature Review} 
\author{
    \IEEEauthorblockN{Simon Schneider}
    \IEEEauthorblockA{
    Hamburg University of Technology\\
    Hamburg, Germany}
    \and
    \IEEEauthorblockN{Ananya Saha}
    \IEEEauthorblockA{
    Hamburg University of Technology\\
    Hamburg, Germany}
    \and
    \IEEEauthorblockN{Emanuele Mezzi}
    \IEEEauthorblockA{
    Vrije Universiteit Amsterdam\\
    Amsterdam, The Netherlands}
    \and
    \IEEEauthorblockN{Katja Tuma}
    \IEEEauthorblockA{
    Vrije Universiteit Amsterdam\\
    Amsterdam, The Netherlands}
    \and
    \IEEEauthorblockN{Riccardo Scandariato}
    \IEEEauthorblockA{
    Hamburg University of Technology\\
    Hamburg, Germany}
}

\maketitle

\begin{abstract}
AI-based systems leverage recent advances in the field of AI/ML by combining traditional software systems with AI components.
Applications are increasingly being developed in this way.
Software engineers can usually rely on a plethora of supporting information on how to use and implement any given technology.
For AI-based systems, however, such information is scarce.
Specifically, guidance on how to securely design the architecture is not available to the extent as for other systems.

We present 16 architectural security guidelines for the design of AI-based systems that were curated via a multi-vocal literature review.
The guidelines could support practitioners with actionable advice on the secure development of AI-based systems.
Further, we mapped the guidelines to typical components of AI-based systems and observed a high coverage where 6 out of 8 generic components have at least one guideline associated to them. 
\end{abstract}


\section{Introduction}
\label{sec:introduction}

With impactful progress in the field of AI/ML, software systems increasingly incorporate these emerging technologies for a wide variety of use-cases~\cite{Amershi19_se_for_ml, Deng18_ai_rising_wave}.
AI-based systems, i.e., software systems containing AI components~\cite{Bosch20_engineering_ai, Martinez-Fernandez22_se_for_ai-based}, face additional security challenges~\cite{Hu21_ai_security} compared to traditional systems.
The architecture of AI-based systems is one dimension of security that needs to be considered during their development.

Typically, software engineers can rely on an abundance of free and readily available information on how to use specific security technology solutions~\cite{Acar17_developrs_need_support}.
For AI-based systems, however, such guidance is limited when it comes to the architectural design, meaning, how AI components can be integrated into a software system's architecture securely.
Resources such as best-practice recommendations, architectural design patterns, or similar could support practitioners in creating more secure systems but are not yet available to the same extent as for other types of systems.

Others identified the lack of such guidance. 
Mucchini and Vaidhyanathan~\cite{Mucchini21_architecture_for_ml_based} have discussed open challenges related to the architecture design process of AI-based systems.
Lewis et al.~\cite{Lewis21_architecture_challenges_for_ml} mention the creation of software architecture practices and architecture patterns as open challenges that would increase the security and adoption of AI-based systems.

In this paper, we present architectural security guidelines for AI-based systems.
We conducted a multi-vocal literature review (MLR) of the relevant academic and gray literature, resulting in the curation of 16 guidelines.
For the gray literature study, we looked for relevant resources (i.e., best-practices, design guidelines, implementation guidance, etc.) published by organizations that are typically consulted by practitioners.
For the systematic literature review (SLR) we identified relevant academic publications.

We address the following research questions (RQs):

\question{\hspace{-0.5mm}\textbf{1}: What is the actionable guidance in the literature with regards to designing secure AI-based systems?}

\noindent
We aimed to collect support provided for practitioners when designing the architecture of AI-based systems, focusing on security.
To this end, we conducted a gray literature study and an SLR to identify existing work on the topic.

\question{\hspace{-0.5mm}\textbf{1.1}: What guidelines are suggested and from which sources?}

\noindent
Actionable, technical guidelines for software development activities are a valuable asset to enhance the security of software systems.
We analyzed the results from the MLR and distilled guidelines for the architectural design of AI-based systems.

\question{\hspace{-0.5mm}\textbf{1.2}: What rationales are given as justification for introducing each guideline?}

\noindent
The reasons for introducing recommendations can be diverse.
To identify what drives the formulation of such recommendations, we examined the rationales provided for each found guideline.
Such results could offer a basis to ensure that no important requirements or stakeholders are neglected.

\question{\hspace{-0.5mm}\textbf{2}: What is the coverage of the guidelines with regards to the typical components of AI-based software systems?}

\noindent
We mapped the presented guidelines to the important components of AI-based systems adapted from Kästner~\cite{Kastner22_ml_in_production}.
We identified those components that are already addressed by the current research and those that are neglected so far.

\section{Systematic Review of Academic Literature}
\label{sec:slr}

\begin{figure}
    \centering
    \includegraphics[width = 0.9\linewidth]{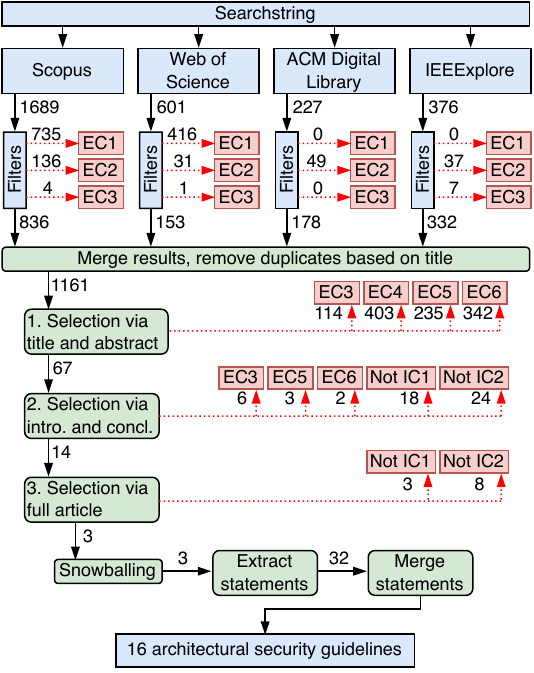}
    \vspace{-2mm}
    \caption{Methodology of the conducted SLR and extraction of guidelines.}
    \vspace{-4mm}
    \label{fig:methodology}
\end{figure}

Figure~\ref{fig:methodology} shows the methodology we applied for the SLR.
We followed established recommendations for conducting systematic reviews~\cite{Kitchenham04_procedures, Kitchenham07_guidelines, Ralph20_empirical_standards} to design the methodology.
A replication package with artifacts for all steps is  available~\cite{replication_package}.

\subsection{Selection of Papers}
\label{sub:paper_selection}

\emph{Scientific Databases:}
We considered four formal scientific databases as sources for the collection of papers, \textit{Scopus},~\footnote{https://www.scopus.com} \textit{Web of Science},~\footnote{https://www.webofscience.com/wos/} \textit{ACM Digital Library},~\footnote{https://dl.acm.org/} and \textit{IEEExplore}.~\footnote{https://ieeexplore.ieee.org/Xplore}
They form a comprehensive set covering the majority of relevant research papers in the domain of computer science. 
We had initially also examined results on \textit{Google Scholar~\footnote{https://scholar.google.de/}}, but found no additional relevant ones compared to the formal databases.

\emph{Search String:}
To construct an appropriate search string, the two first authors independently performed explorative searches in the databases (following the ``extraordinary attributes'' suggestion for systematic reviews presented by Ralph et al.~\cite{Ralph20_empirical_standards}).
They used search terms related to the topic and iteratively created a list of relevant keywords, synonyms, alternative spellings, etc. based on the findings~\cite{Kitchenham07_guidelines}.
Both authors adapted their search strings in approximately 10 iterations and by looking at roughly 100 different papers. 
Afterwards, they created the final search string together, combining their results from the explorative searches.
The final search string combines different keywords for the scope (AI-based systems) and different keywords for the topic (guidelines):

\begin{center}
\small
\ttfamily
("ai system" OR "ai component" OR "ai enabled" OR "ai based" OR "ai empowered" OR "ai augmented") AND (guideline* OR "best practices" OR "best-practices" OR smell* OR rule* OR antipattern* OR "design principles") 

\end{center}

\noindent 
It was adapted for the searches to fit the databases' formatting requirements where necessary.
Figure~\ref{fig:methodology} shows the number of search results for each of the four databases.

\emph{In- / Exclusion Criteria:}
The inclusion and exclusion criteria (IC and EC) were constructed based on the formulated research questions. 
During the explorative searches preceding the creation of the search string, we noticed that the search results would inevitably contain many entries not relevant for our study.
This is mainly due to two reasons: (1) the domain (computer science) could not be reliably enforced with the search parameters and because of the use of AI technologies in diverse fields many results from other research areas come up; and (2) the scope (AI-based systems) is hard to enforce with the search string and many results come up that concern the AI models and algorithms themselves instead of systems that use them.
The in- and exclusion criteria address this.

\noindent
\textbf{Inclusion criteria:}
\begin{description}
    \item[IC1:] The paper presents guidelines for AI-based systems.
    \item[IC2:] The guidelines concern the systems' architecture.
\end{description}

\noindent
\textbf{Exclusion criteria:}
\begin{description}
    \item[EC1:] The paper is not in the subject area of computer science.
    \item[EC2:] The paper was not published in the last ten years.
    \item[EC3:] The paper was not published in a conference, journal, or book.
    \item[EC4:] The paper reports on a specific use-case of AI models.
    \item[EC5:] The paper considers only specific industries without generalization, e.g., the automotive or aviation industry.
    \item[EC6:] The paper is not related to AI-based systems but, e.g., the AI models themselves or non-AI topics.
\end{description}
\noindent 
EC1 -- EC3 were realized with the databases' search filters where possible and subsequently checked for in the manual selection as well.
The search queries per database that include these filters can be found in the replication package~\cite{replication_package}.

\emph{Selection Process:}
The results of the searches in the four databases were merged by removing duplicates based on the articles' titles.
This resulted in a total of 1161 articles.
The selection was conducted in three phases: based on (1) title and abstract; (2) introduction and conclusion section; (3) the complete content.
This is an established process and reduces the required effort.
We aimed to ensure that no relevant articles were excluded by applying a liberal inclusion tendency and only excluding articles that are clearly not relevant to the study.

For each phase, two authors independently applied the criteria to each article.
We excluded articles that meet any EC in phases 1 and 2 and articles that do not meet both IC in phase 3.
A third author checked conflicts between the first two decisions and made a final decision.
Note, that we only assigned a single EC even though multiple could apply to a paper. 
For each paper, we went through the list of ECs in order and assigned it to the first one that applied to the paper, if any.
Although EC3 was meant to be enforced with search parameters of the databases, a total of 120 results were still returned by the searches and subsequently excluded manually.

For each phase of the SLR, we calculated the Cohen's Kappa~\cite{Vieira10_cohens_kappa} to measure the agreement between the two authors.
We observed a substantial agreement in phase 1 (Cohen's Kappa of 0.70), almost perfect agreement in phase 2 (0.81), and perfect agreement in phase 3 (1.0) (classification according to Landis and Koch~\cite{Landis77_agreement}).
The extraction of statements showed a 76\% agreement.
The observed agreement is a positive indicator for the validity of the selection process.
A third author as mediator to solve conflicts further strengthens this.

\subsection{Curation of Architectural Security Guidelines}

\begin{table}
\centering
    \caption{Filtering criteria for statements in the analysed articles.}
    \label{tbl:filtering_criteria}
    \begin{tabular}{cm{6.5cm}} 
        \toprule
        \emph{Concrete} & The statement clearly describes a (non-) desired system property or pattern. \\
        \midrule
        \emph{In Scope} & The statement refers to the architectural design or architectural components of AI-based systems. \\
        \bottomrule
    \end{tabular}
\end{table}

We analyzed the selected articles and extracted statements that are relevant for answering our formulated research questions.
Specifically, we looked for concrete descriptions of a desired or non-desired property of AI-based systems that refers to their architectural components. 
Table~\ref{tbl:filtering_criteria} presents these filtering criteria we applied to identify relevant statements.
They were adopted from a comparable study about architectural security rules for microservice applications~\cite{BamboreTukaram22_security_benchmark_microservices}.

As for the selection of articles, two authors independently performed this extraction of statements and a third author helped solve conflicts in a discussion.
The retained statements were then discussed with all authors to create a comprehensive list of guidelines that represent all statements found in the analyzed articles.
In this discussion, we constructed guidelines out of the found statements by merging similar ones and formulating them in a similar style. 
As a final check, all three authors validated that the final list of guidelines represents all statements extracted from the selected articles.

\section{Gray Literature Study with Negative Results}
\label{sec:gray_lit_study}

\begin{figure}
    \centering
    \includegraphics[width = 0.8\linewidth]{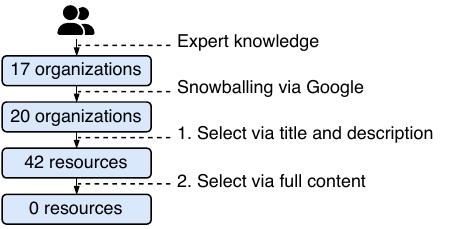}
    \vspace{-2mm}
    \caption{Methodology of gray literature study with negative results.}
    \vspace{-4mm}
    \label{fig:methodology_gray_literature}
\end{figure}

We had initially planned to curate architectural security guidelines via an MLR~\cite{Garousi19_guidelines_mvlr}.
However, when we conducted the gray literature study, it did not yield any relevant resources.

\begin{table}
    \centering
	\caption{Organizations considered in the gray literature study.\vspace{-2mm}} 
	\label{tbl:organizations}
    \begin{tabular}{lclc}
        \toprule
        \textbf{Organization} & \textbf{Resources} & \textbf{Organization} & \textbf{Resources} \\
         & \textbf{after phase 1} & & \textbf{after phase 1} \\
        \midrule
        BSA & 11 & MITRE & 10\\
        BSI & 7 & MITRE ATLAS & \\
        CERT-EU &  & NCSC & 2 \\
        CIS &  & NIST & 2  \\
        CISA &  & NSA  &  \\
        CSA & 1 & OECD & 2 \\
        ENISA & 7 & OpenSSF & \\
        ESI &  & OWASP &  \\
        ESCO &  & SANS Institute &  \\
        FIRST &  & SAFECode & \\
        
        \midrule 
        \multicolumn{4}{l}{\textbf{Total: 42}} \\
        \bottomrule
    \end{tabular}
    \vspace{-5mm}
\end{table}

Figure~\ref{fig:methodology_gray_literature} shows the methodology we followed for the gray literature study.
It is adapted from the one used by Tukaram et al.~\cite{BamboreTukaram22_security_benchmark_microservices}.
We looked for resources published by large organizations that offer best-practice guides and other resources for developers.
We created a list of 17 such organizations via the expert knowledge of six security researchers, three of which are authors and the rest are colleagues of them.
The list was extended by forward snowballing via the Google Search engine.
Table~\ref{tbl:organizations} shows the resulting list of 20 organizations.
Two authors went through all resources published by each of the organizations to select all those that are related to AI and of which the title and description indicated that they could contain information about the architectural design of AI-based systems.
This resulted in the selection of 42 resources.

They then independently checked those resources in detail for adherence to the formulated in- / exclusion criteria (as presented above in Section~\ref{sec:slr}).
Upon further inspection, both authors decided that no resource fit the inclusion criteria.
None of the found resources presents guidelines or any other architectural information for AI-based systems.

The examined organizations are highly regarded by practitioners and often serve as a source of the latest, high-quality information on how to implement and use specific technological solutions.
Considering this, the negative outcome of this study is worrying, as it indicates a lack of available guidance for developers working on AI-based systems.
This can in turn lead to architectural security issues of those systems.

\section{Guidelines Curated from Literature (RQ1)}
\label{sec:results}

\begin{table*}
    \centering
	\caption{Primary Studies.\vspace{-2mm}}
	\label{tbl:selected_papers}
    \begin{tabular}{cllc}
        \toprule
        \textbf{ID} & \textbf{Title} & \textbf{Authors} & \textbf{Year} \\
        \midrule
        S1 & Responsible AI Pattern Catalogue: A Collection of Best Practices for AI Governance and Engineering~\cite{Lu24_P1_responsible_ai} & Lu et al. & 2023 \\
        S2 & Characterizing Technical Debt and Antipatterns in AI-Based Systems: A Systematic Mapping Study~\cite{Bogner21_P2_technical_debt} & Bogner et al. & 2021 \\
        S3 & A Survey of Privacy Risks and Mitigation Strategies in the Artificial Intelligence Life Cycle~\cite{Shahriar23_P3_privacy_risks} & Shahriar et al. & 2023 \\
        
        \bottomrule
    \end{tabular}
    \vspace{-5mm}
\end{table*}

The conducted SLR resulted in the selection of three articles that present architectural guidelines for designing AI-based systems.
They are presented as studies S1 -- S3 in Table~\ref{tbl:selected_papers}.

Study S1~\cite{Lu24_P1_responsible_ai} presents the findings of a multi-vocal literature review on design patterns for AI-based systems that foster responsible AI.
Many of the presented patterns are implementation-oriented, but procedural and other recommendations not relevant for us are also contained.
Study S2~\cite{Bogner21_P2_technical_debt} lists types of technical debt and antipatterns that apply to AI-based systems and were identified via a systematic mapping study.
Specific instances of the proposed catalog of technical debt and antipatterns are only occasionally presented, and the accompanying replication package also lacks details in some cases.
We thus also investigated the mentioned primary studies of S2 for more details where needed.
Study S3~\cite{Shahriar23_P3_privacy_risks} presents a survey on privacy risks of AI-based systems along with possible mitigations.
Due to its focus on privacy, most presented risks are related to the collection and use of data.

Although we had not set out to conduct a tertiary study, the selection process produced only secondary studies.
We hypothesize that this happened because we looked for papers presenting generalized results instead of findings of specific case-studies.
We deduce, that the literature on the matter seems to be still in its infancy.
A look at the primary studies used in S1 -- S3 confirms this observation. 
The majority of them describe observations of specific cases.

\begin{table*}
    \centering
	\caption{Architectural security guidelines for AI-based systems (RQ1.1) and the rationale given for their introduction (RQ1.2). \\A filled box ($\blacksquare$) under ``S1 S2 S3'' indicates, that the guideline is mentioned in the corresponding source S1 -- S3.\vspace{-2mm}}
	\label{tbl:guidelines}
    \begin{tabular}{P{0.3cm}p{12cm}P{1.3cm}p{2.8cm}}
        \toprule
        \textbf{ID} & \textbf{Guideline} & \textbf{S1 S2 S3} & \textbf{Rationale}\\
        \midrule
        G1 & \textbf{AI-mode switcher:} An AI mode switcher can be used to let the human user control whether AI components' outcomes are followed automatically or taken as suggestions and reviewed by the user. & $\blacksquare  \; \square  \; \square $ & Safety, security \\
        G2 & \textbf{Multi-model decision maker:} Employing a multi-model decision maker consisting of more than one different AI components can increase the system's reliability and fault-tolerance. & $\blacksquare  \; \square  \; \square $ & Ethics, performance, \mbox{reliability} \\
        G3 & \textbf{Redundant AI components:} Deploying multiple identical AI components in parallel can ensure tolerance towards a malicious or otherwise disfunctioning AI component. & $\blacksquare  \; \square  \; \square $ & Ethics, reliability, \mbox{security} \\
        G4 & \textbf{Ethical black box:} AI components' outcomes and related system information should be recorded in a black box. If rewards are used as incentive for ethical behaviour, they should be recorded as well.  & $\blacksquare  \; \square  \; \square $ & Ethics, explainability, responsible AI \\
        G5 & \textbf{Continuous monitoring:} AI components' outcomes should continuously be checked for their adherence to ethical requirements. & $\blacksquare  \; \square  \; \square $ & Ethics \\
        G6 & \textbf{Monitoring AI components' discrepancies:} If multiple redundant AI components are deployed, their outcomes should be monitored to identify discrepancies. & $\blacksquare  \; \square  \; \square $ & Accountability, safety, security \\
        G7 & \textbf{Adapting traditional design methods:} Architectural design methods used for traditional systems (e.g., UML diagrams) can be adapted and used for AI-based systems as well and can model ethical requirements. & $\blacksquare  \; \square  \; \square $ & Ethics, Responsible AI \\
        G8 & \textbf{Access via APIs:} AI-based systems can be made available via APIs instead of allowing users to run them locally to keep control over their use. & $\blacksquare  \; \square  \; \square $ & \mbox{Responsible AI}, \mbox{Security} \\
        G9 & \textbf{Bill of materials registry:} A bill of materials registry should be integrated that keeps a machine-readable record of all used components and their properties. & $\blacksquare  \; \square  \; \square $ & Ethics, Responsible AI, Security \\
        G10 & \textbf{Reusable components:} Components should be implemented as reusable units with clearly defined boundaries and explicit interfaces. Complex dependencies and entanglement should be avoided.  & $\square  \; \blacksquare  \; \square $ & Maintainability \\
        G11 & \textbf{Compatible data types:} Incompatible data types between AI and other components should be avoided. & $\square  \; \blacksquare  \; \square $ & N/A \\
        G12 & \textbf{Single programming language:} The use of multiple programming languages in the same system should be avoided where possible. & $\square  \; \blacksquare  \; \square $ & Maintainability, security \\
        G13 & \textbf{No undeclared consumers:} Undeclared consumers of AI components should be avoided and access control for the outcomes of the AI components should be enforced. & $\square  \; \blacksquare  \; \square $ & Security \\
        G14 & \textbf{Performance monitoring:} There should be a component that monitors the performance of the AI components concerning their prediction accuracy. & $\square  \; \blacksquare  \; \square $ & Maintainability \\
        G15 & \textbf{Sufficient resources:} Sufficient computing and memory resources have to be available for the AI components. & $\square  \; \blacksquare  \; \blacksquare $ & Performance \\
        G16 & \textbf{Secure data storage:} When data is collected for re-training, it has to be stored securely and according to applicable regulations such as the GDPR. This includes the use of encryption techniques, a firewall between data storage and network, and enforcement of access control. & $\square  \; \square  \; \blacksquare $ & Compliance, privacy, \mbox{security} \\
        
        \bottomrule
    \end{tabular}
    \vspace{-5mm}
\end{table*}

The analysis of the identified articles resulted in the formulation of 16 architectural security guidelines for AI-based systems.
Table~\ref{tbl:guidelines} presents them as G1 -- G16.
They are described briefly in the following.
\textbf{G1} suggests adding an AI-mode switcher to the architecture that acts ``like a kill switch''~\cite{Lu24_P1_responsible_ai} for the AI component(s).
The switch changes the mode of the system from automatically acting on outcomes of the AI component(s) to treating them as suggestions that have to be reviewed by the user first.
Deploying multiple different (\textbf{G2}) or identical (\textbf{G3}) AI components in parallel can be advantageous for the case that one of them acts maliciously or otherwise dysfunctional and can increase the system's reliability by employing a consensus mechanism between them~\cite{Nafreen20_architecture-based_software_reliability}.
The outcomes of the AI component(s) and incentives given for them should be recorded in a ``black box''~\cite{Winfield18_ethical_blackbox} as known from airplanes to enable an analysis of what caused an undesired system behavior in the case that this happens (\textbf{G4}).
The outcomes should also be continuously checked against ethical requirements to ensure they are still met even after re-training the models (\textbf{G5})~\cite{Staples16_continuous_validation}, and they should be compared against each other to identify discrepancies between the models' outcomes (\textbf{G6})~\cite{Miguel21_accountability_of_ai}.
\textbf{G7} suggest adapting existing tools and techniques for designing the architecture of software systems to AI-based systems as well.
Further, opening the access to AI-based systems to users can be done via APIs instead of letting users deploy them on their own machines (\textbf{G8}).
This keeps the developers in control over the system and its acceptable use.
\textbf{G9} suggests keeping a bill of materials registry that records relevant information about all components that are part of the AI-based system and the data used for training the AI component(s).
This allows traceability and transparency, especially when issues in the systems' behavior occur.
\textbf{G10} addresses the observation that the code of AI-based systems tends to be poorly structured and not follow established coding practices such as well-defined interfaces of components~\cite{Sculley15_hidden_technical_debt, Saleema19_se_for_ml}.
\textbf{G11} also concerns the connection between AI component(s) and others, noting that incompatible data types can often be an issue leading to a large amount of glue code and should therefore be avoided~\cite{Lewis19_component_mismatch}.
According to \textbf{G12}, the use of multiple programming languages in the development of an AI-based system should be avoided where possible, because it makes testing, automated refactoring, and deployment more difficult~\cite{Schelter15_challenges_ml_management}.
\textbf{G13} states that access to the outcomes of the AI component(s) should be secured by enforcing access control and identifying all consumers~\cite{Sculley15_hidden_technical_debt}.
Otherwise, control over potential misuses of the outcomes can not reliably be kept.
\textbf{G14} states that a monitoring component should be deployed to identify performance degradation in the prediction accuracy of the AI component(s)~\cite{Agarwal16_contextual_decisions}.
\textbf{G15} addresses the issue that AI-based systems are often deployed with insufficient resources for the AI component to run properly when moving from development to production~\cite{Lewis19_component_mismatch}.
When data is collected by the deployed AI-based system for re-training purposes, it has to be protected in accordance to applicable regulations and using traditional security mechanisms (\textbf{G16})~\cite{Jain16_big_data_privacy, Kasneci23_chatgpr_for_good}.

As visible in the column ``prevalence'' in Table~\ref{tbl:guidelines}, G15 is the only guideline that was mentioned in more than one of the sources S1 -- S3.
A bigger overlap between the sources would be desirable, as it would indicate their validity.
Most likely, this is due to us only identifying three sources for architectural guidelines, which additionally approach the issue from different motivations and are secondary studies.
Therefore, we hope for more related work in the future that would allow assessing overlaps between sources.

We summarize that, overall, research on actionable support for practitioners working on the architecture of AI-based systems is ongoing and that various guidelines can be found.
On the other hand, the small number of identified articles and rather small number of resulting guidelines shows the limited extent of tangible guidance.

\answer{\hspace{-0.5mm}\textbf{1.1}: We identified 16 architectural security guidelines in the academic literature, presented in Table~\ref{tbl:guidelines}.
As described in Section~\ref{sec:gray_lit_study}, no guidelines were found in the gray literature we examined.}

Table~\ref{tbl:guidelines} presents in the right-most column the rationales given as justification for introducing each guideline.
For most guidelines, it reflects the general motivation of the corresponding source.
For example, seven out of nine guidelines found in S1 have ``ethics'' or ``responsible AI'' as one of the rationales, i.e., the requirement that the AI-based systems behave in an ethical and responsible way, as is expected by the paper's title and goal.
Nevertheless, the rationales are diverse across all guidelines, with 11 distinct ones mentioned, covering different aspects of security, ethics, and general system qualities.

Interestingly, just over half of the guidelines (7) mention more than one individual rationales as justification for their introduction. 
This speaks for the importance of the guidelines, as they seem to support cross-cutting improvement of different quality aspects of the systems.
For some of those guidelines only mentioning one rationale, they are likely to also contribute to further qualities of the AI-based systems. 
For example, the paper proposing G10 mentions maintainability as rationale, however, the systems' security would likely also benefit from clearly defined components as this helps developers gain and maintain an overview of the architecture and prevent security issues.

\answer{\hspace{-0.5mm}\textbf{1.2}: Table~\ref{tbl:guidelines} presents the rationales given for introducing each of the curated guidelines.
A total of 11 individual rationales are mentioned.
They show various aspects of security, ethics, and general system qualities, with ``ethics'' and ``security'' being the most prevalent ones.}

\section{Mapping to Typical Components of AI-based Systems (RQ2)}
\label{sec:mapping}

We mapped the guidelines identified via our SLR to typical components of AI-based systems to examine the coverage of the architecture, i.e., those parts for which guidelines could be found in the literature.
As the basis for the mapping, we took an example architecture proposed by Kästner~\cite{Kastner22_ml_in_production} and generalized those components that were use-case specific. 
It is taken from a popular course on developing AI-based systems and the author is a known expert in the field.
We are aware that other representations of AI-based systems exist and could be used as the basis for the mapping. 
For example, Sculley et al.~\cite{Sculley15_hidden_technical_debt} presented a generic architecture of AI-based systems that includes the production pipeline and detailed data science pipeline used to create the AI components.
In this work, however, we focus on the architecture of the final product and hence selected the architecture by Kästner~\cite{Kastner22_ml_in_production} as the basis for our mapping.
Specifically, G7 is the only identified guideline that is clearly and only related to the development process.

We mapped the guidelines in Table~\ref{tbl:guidelines} to the resulting generic architecture by identifying which component of the architecture each guideline applies to.
The mapping was performed in a discussion with two authors.
No ambiguities occurred, and the connections between guidelines and components are rather straightforward.

\begin{figure}
    \centering
    \includegraphics[width = \linewidth]{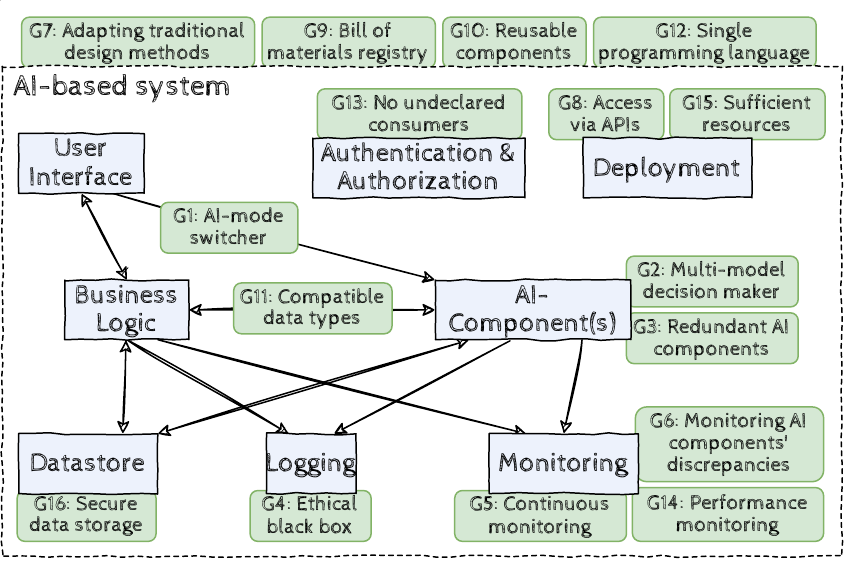}
    \vspace{-5mm}
    \caption{Mapping of architectural security guidelines to typical components of AI-based systems, adapted from Kästner~\cite{Kastner22_ml_in_production}.}
    \label{fig:architecture_mapping}
    \vspace{-4mm}
\end{figure}

Figure~\ref{fig:architecture_mapping} shows the generic architecture and the mapping of the guidelines.
The guidelines are distributed well across the components, and we see a high coverage of the typical components of AI-based systems.
The ``User Interface'' and ``Business Logic'' are the only two that do not have any guidelines directly associated to them, but both have a guideline concerning their interaction with the AI component(s).
Most components only have a single guideline associated to them.

The lack of guidelines for the ``Business Logic'' is surprising.
We would have expected to find recommendations about how to securely handle AI components' outcomes and how the systems' core functionality should address the non-determinism introduced by it.
The comparatively large number of guidelines related to monitoring and logging is based on the prevalence of ethical considerations as rationale for the guidelines.
It also shows a tendency to supervise AI components well to address possibly malicious ones and changing behavior based on re-training.

The high coverage of the typical components of AI-based systems is generally seen as a positive sign for the state of the art in research on the topic.
Most components contained in AI-based systems are considered by at least one guideline. 
We note that different components for the generic architectures could be proposed by others.
We do not claim completeness or generality for either the used architecture by Kästner~\cite{Kastner22_ml_in_production} or our derived one.
However, we believe it to be generic enough to allow the presented general observations and discussion.

\answer{\hspace{-0.5mm}\textbf{2}: Mapping the guidelines to the typical components of AI-based systems showed a high coverage of the components. 
``User Interface'' and ``Business Logic'' are the only two components without a guideline directly associated to them.
Figure~\ref{fig:architecture_mapping} shows the complete mapping.}

\section{Threats to Validity}
\label{sec:threats_to_validity}

All manual steps of the SLR and the gray literature review, the curation of the guidelines, and their mapping to the typical components of AI-based systems could have been subject to selection bias, search bias, extraction errors, and synthesis bias.
This can affect the reliability and generalizability of the presented results.
We addressed this with robust and established methods by performing all steps with two authors independently, solving conflicts with a third author, formulating reproducible in- / exclusion and filtering criteria, measuring agreement with Cohen's Kappa, and publishing a replication package~\cite{replication_package}.

The small number of found articles, which in addition are all secondary studies, suggests that the exclusion criteria might have been too restrictive.
However, all criteria were necessary to exclude a large amount of irrelevant articles and all are well motivated and cater to the research questions.
We believe a lack of research on the topic to be the reason for the small number of selected articles.

The architecture used for the mapping is not based on a sound study but on experience of the author of the initial architecture.
No other architecture was found in the related literature that would lend itself to such a mapping, and we do not claim generality.

\section{Related Work}
\label{sec:related_work}

Many experience reports and case-studies have been published in the related literature that present specific challenges and possible solutions for the architectural design of AI-based systems.
They are covered by the articles that we identified with our SLR and are thus not presented additionally here.

Some authors also presented studies that summarize such specific work into more general sets of recommendations and guidelines, focusing on different aspects of the development of AI-based systems, such as collaboration challenges between practitioners~\cite{Nahar22_collaboration_challenges}, architectural patterns for federated learning systems~\cite{Lo22_architectural_patterns_fls}, specific technical debt~\cite{Sculley15_hidden_technical_debt}, or general challenges of engineering such systems~\cite{Lwakatare19_se_challenges_ml_systems}.
The studies identified with our SLR also fit into this category (presented below).

Tukaram et al.~\cite{BamboreTukaram22_security_benchmark_microservices} curated architectural security rules for microservice applications. 
They analyzed resources published by organizations as we aimed to do with our gray literature study that yielded no results (presented below in Section~\ref{sec:gray_lit_study}).
Except for the different domain, their paper and results are similar to ours, and we adapted parts of their methodology.

To the best of our knowledge, no work has been presented in the academic literature that aimed to curate architectural security guidelines for AI-based systems.

\section{Research Outlook}
\label{sec:research_outlook}

\begin{enumerate}[leftmargin=*]

    \item More research efforts should be applied to the architectural security of AI-based systems, as it seems to trail behind the work on other security aspects. 
    In particular, we are astonished that the large, emerging literature on attacks to AI/ML has not translated yet to advice related to possible countermeasures.
    \item Architectural security guidelines for AI-based systems would greatly benefit from the expertise of the practitioners at the forefront of their development. In this respect, a large-scale survey could be a useful instrument.
    \item Attention should be given to the problem of compliance and certification with respect to these (and future) design guidelines. Methodologies and tools should be created that allow analyses of these aspects. 
\end{enumerate}

\section*{Acknowledgement}
\noindent This work was supported by EU-funded project Sec4AI4Sec (grant no. 101120393).

\balance
\bibliographystyle{ieeetr}
\bibliography{ai_guidelines}

\begin{thebibliography}{10}

\bibitem{Amershi19_se_for_ml}
S.~Amershi, A.~Begel, C.~Bird, R.~DeLine, H.~Gall, E.~Kamar, N.~Nagappan,
  B.~Nushi, and T.~Zimmermann, ``Software engineering for machine learning: A
  case study,'' in {\em 2019 IEEE/ACM 41st International Conference on Software
  Engineering: Software Engineering in Practice (ICSE-SEIP)}, pp.~291--300,
  2019.

\bibitem{Deng18_ai_rising_wave}
L.~Deng, ``Artificial intelligence in the rising wave of deep learning: The
  historical path and future outlook [perspectives],'' {\em IEEE Signal
  Processing Magazine}, vol.~35, no.~1, pp.~180--177, 2018.

\bibitem{Bosch20_engineering_ai}
J.~Bosch, I.~Crnkovic, and H.~H. Olsson, ``Engineering ai systems: A research
  agenda,'' 2020.

\bibitem{Martinez-Fernandez22_se_for_ai-based}
S.~Mart\'{\i}nez-Fern\'{a}ndez, J.~Bogner, X.~Franch, M.~Oriol, J.~Siebert,
  A.~Trendowicz, A.~M. Vollmer, and S.~Wagner, ``Software engineering for
  ai-based systems: A survey,'' {\em ACM Trans. Softw. Eng. Methodol.},
  vol.~31, apr 2022.

\bibitem{Hu21_ai_security}
Y.~Hu, W.~Kuang, Z.~Qin, K.~Li, J.~Zhang, Y.~Gao, W.~Li, and K.~Li,
  ``Artificial intelligence security: Threats and countermeasures,'' {\em ACM
  Comput. Surv.}, vol.~55, nov 2021.

\bibitem{Acar17_developrs_need_support}
Y.~Acar, C.~Stransky, D.~Wermke, C.~Weir, M.~L. Mazurek, and S.~Fahl,
  ``Developers need support, too: A survey of security advice for software
  developers,'' in {\em 2017 IEEE Cybersecurity Development (SecDev)},
  pp.~22--26, 2017.

\bibitem{Mucchini21_architecture_for_ml_based}
H.~Muccini and K.~Vaidhyanathan, ``Software architecture for ml-based systems:
  What exists and what lies ahead,'' in {\em 2021 IEEE/ACM 1st Workshop on AI
  Engineering - Software Engineering for AI (WAIN)}, (Los Alamitos, CA, USA),
  pp.~121--128, IEEE Computer Society, may 2021.

\bibitem{Lewis21_architecture_challenges_for_ml}
G.~A. Lewis, I.~Ozkaya, and X.~Xu, ``Software architecture challenges for ml
  systems,'' in {\em 2021 IEEE International Conference on Software Maintenance
  and Evolution (ICSME)}, pp.~634--638, 2021.

\bibitem{Kastner22_ml_in_production}
C.~K{\"a}stner, ``{Machine Learning in Production: from Models to Products},''
  2022.

\bibitem{Kitchenham04_procedures}
B.~Kitchenham, ``Procedures for performing systematic reviews,'' {\em Keele,
  UK, Keele Univ.}, vol.~33, 08 2004.

\bibitem{Kitchenham07_guidelines}
B.~Kitchenham and S.~Charters, ``Guidelines for performing systematic
  literature reviews in software engineering,'' vol.~2, 01 2007.

\bibitem{Ralph20_empirical_standards}
P.~Ralph, N.~b. Ali, S.~Baltes, D.~Bianculli, J.~Diaz, Y.~Dittrich, N.~Ernst,
  M.~Felderer, R.~Feldt, A.~Filieri, {\em et~al.}, ``Empirical standards for
  software engineering research,'' {\em arXiv preprint arXiv:2010.03525}, 2020.

\bibitem{replication_package}
S.~Schneider, A.~Saha, E.~Mezzi, K.~Tuma, and R.~Scandariato, ``{Replication
  package for 'Architectural Security Guidelines for AI-based Systems'},'' May
  2024.

\bibitem{Vieira10_cohens_kappa}
S.~M. Vieira, U.~Kaymak, and J.~M.~C. Sousa, ``Cohen's kappa coefficient as a
  performance measure for feature selection,'' in {\em International Conference
  on Fuzzy Systems}, pp.~1--8, 2010.

\bibitem{Landis77_agreement}
J.~R. Landis and G.~G. Koch, ``The measurement of observer agreement for
  categorical data,'' {\em Biometrics}, vol.~33, no.~1, pp.~159--174, 1977.

\bibitem{BamboreTukaram22_security_benchmark_microservices}
A.~Bambhore~Tukaram, S.~Schneider, N.~E. D\'{\i}az~Ferreyra, G.~Simhandl,
  U.~Zdun, and R.~Scandariato, ``Towards a security benchmark for the
  architectural design of microservice applications,'' in {\em Proceedings of
  the 17th International Conference on Availability, Reliability and Security},
  ARES '22, (New York, NY, USA), Association for Computing Machinery, 2022.

\bibitem{Garousi19_guidelines_mvlr}
V.~Garousi, M.~Felderer, and M.~V. Mäntylä, ``Guidelines for including grey
  literature and conducting multivocal literature reviews in software
  engineering,'' {\em Information and Software Technology}, vol.~106,
  pp.~101--121, 2019.

\bibitem{Lu24_P1_responsible_ai}
Q.~Lu, L.~Zhu, X.~Xu, J.~Whittle, D.~Zowghi, and A.~Jacquet, ``Responsible ai
  pattern catalogue: A collection of best practices for ai governance and
  engineering,'' {\em ACM Comput. Surv.}, vol.~56, apr 2024.

\bibitem{Bogner21_P2_technical_debt}
J.~Bogner, R.~Verdecchia, and I.~Gerostathopoulos, ``Characterizing technical
  debt and antipatterns in ai-based systems: A systematic mapping study,'' in
  {\em 2021 IEEE/ACM International Conference on Technical Debt (TechDebt)},
  pp.~64--73, IEEE, May 2021.

\bibitem{Shahriar23_P3_privacy_risks}
S.~Shahriar, S.~Allana, S.~M. Hazratifard, and R.~Dara, ``A survey of privacy
  risks and mitigation strategies in the artificial intelligence life cycle,''
  {\em IEEE Access}, vol.~11, pp.~61829--61854, 2023.

\bibitem{Nafreen20_architecture-based_software_reliability}
M.~Nafreen, S.~Bhattacharya, and L.~Fiondella, ``Architecture-based software
  reliability incorporating fault tolerant machine learning,'' in {\em 2020
  Annual Reliability and Maintainability Symposium (RAMS)}, pp.~1--6, 2020.

\bibitem{Winfield18_ethical_blackbox}
A.~F. Winfield and M.~Jirotka, ``The case for an ethical black box,'' {\em In:
  Gao Y., Fallah S., Jin Y., Lekakou C. (eds) Towards Autonomous Robotic
  Systems. TAROS 2017. Lecture Notes in Computer Science, vol 10454. Springer,
  Cham}, 2018.

\bibitem{Staples16_continuous_validation}
M.~Staples, L.~Zhu, and J.~Grundy, ``Continuous validation for data analytics
  systems,'' in {\em 2016 IEEE/ACM 38th International Conference on Software
  Engineering Companion (ICSE-C)}, pp.~769--772, 2016.

\bibitem{Miguel21_accountability_of_ai}
B.~S. Miguel, A.~Naseer, and H.~Inakoshi, ``Putting accountability of ai
  systems into practice,'' in {\em Proceedings of the Twenty-Ninth
  International Joint Conference on Artificial Intelligence}, IJCAI'20, 2021.

\bibitem{Sculley15_hidden_technical_debt}
D.~Sculley, G.~Holt, D.~Golovin, E.~Davydov, T.~Phillips, D.~Ebner,
  V.~Chaudhary, M.~Young, J.-F. Crespo, and D.~Dennison, ``Hidden technical
  debt in machine learning systems,'' in {\em Advances in Neural Information
  Processing Systems} (C.~Cortes, N.~Lawrence, D.~Lee, M.~Sugiyama, and
  R.~Garnett, eds.), vol.~28, Curran Associates, Inc., 2015.

\bibitem{Saleema19_se_for_ml}
S.~Amershi, A.~Begel, C.~Bird, R.~DeLine, H.~Gall, E.~Kamar, N.~Nagappan,
  B.~Nushi, and T.~Zimmermann, ``Software engineering for machine learning: A
  case study,'' in {\em 2019 IEEE/ACM 41st International Conference on Software
  Engineering: Software Engineering in Practice (ICSE-SEIP)}, pp.~291--300,
  2019.

\bibitem{Lewis19_component_mismatch}
G.~A. Lewis, S.~Bellomo, and A.~Galyardt, ``Component mismatches are a critical
  bottleneck to fielding ai-enabled systems in the public sector,'' in {\em
  AAAI FSS-19: Artificial Intelligence in Government and Public Sector}, 2019.

\bibitem{Schelter15_challenges_ml_management}
S.~Schelter, F.~Biessmann, T.~Januschowski, D.~Salinas, S.~Seufert, and
  G.~Szarvas, ``On challenges in machine learning model management,'' {\em IEEE
  Data Engineering Bulletin}, 2015.

\bibitem{Agarwal16_contextual_decisions}
A.~Agarwal, S.~Bird, M.~Cozowicz, L.~Hoang, J.~Langford, S.~Lee, J.~Li,
  D.~Melamed, G.~Oshri, O.~Ribas, {\em et~al.}, ``Making contextual decisions
  with low technical debt,'' {\em arXiv preprint arXiv:1606.03966}, 2016.

\bibitem{Jain16_big_data_privacy}
P.~Jain, M.~Gyanchandani, and N.~Khare, ``Big data privacy: a technological
  perspective and review,'' {\em Journal of Big Data}, vol.~3, 11 2016.

\bibitem{Kasneci23_chatgpr_for_good}
E.~Kasneci, K.~Sessler, S.~Küchemann, M.~Bannert, D.~Dementieva, F.~Fischer,
  U.~Gasser, G.~Groh, S.~Günnemann, E.~Hüllermeier, S.~Krusche, G.~Kutyniok,
  T.~Michaeli, C.~Nerdel, J.~Pfeffer, O.~Poquet, M.~Sailer, A.~Schmidt,
  T.~Seidel, M.~Stadler, J.~Weller, J.~Kuhn, and G.~Kasneci, ``Chatgpt for
  good? on opportunities and challenges of large language models for
  education,'' {\em Learning and Individual Differences}, vol.~103, p.~102274,
  2023.

\bibitem{Nahar22_collaboration_challenges}
N.~Nahar, S.~Zhou, G.~Lewis, and C.~K\"{a}stner, ``Collaboration challenges in
  building ml-enabled systems: communication, documentation, engineering, and
  process,'' in {\em Proceedings of the 44th International Conference on
  Software Engineering}, ICSE '22, (New York, NY, USA), p.~413–425,
  Association for Computing Machinery, 2022.

\bibitem{Lo22_architectural_patterns_fls}
S.~K. Lo, Q.~Lu, L.~Zhu, H.-Y. Paik, X.~Xu, and C.~Wang, ``Architectural
  patterns for the design of federated learning systems,'' {\em Journal of
  Systems and Software}, vol.~191, p.~111357, 2022.

\bibitem{Lwakatare19_se_challenges_ml_systems}
L.~E. Lwakatare, A.~Raj, J.~Bosch, H.~H. Olsson, and I.~Crnkovic, ``A taxonomy
  of software engineering challenges for machine learning systems: An empirical
  investigation,'' in {\em Agile Processes in Software Engineering and Extreme
  Programming} (P.~Kruchten, S.~Fraser, and F.~Coallier, eds.), (Cham),
  pp.~227--243, Springer International Publishing, 2019.

\end{thebibliography}

\end{document}